%% file: 24ESSERC_MOTHRA.tex
\documentclass[conference,10pt,twoside,twocolumn]{IEEEtran}
\usepackage{cite}
\usepackage[T1]{fontenc}
\usepackage{graphicx}
\usepackage{xcolor}
\usepackage{amssymb}
\usepackage{amsmath}
\usepackage{graphicx}
\usepackage{subfigure}
\usepackage{booktabs} 
\usepackage{algorithm}
\usepackage{algorithmic}
\usepackage{amsthm}
\usepackage{multirow}
\usepackage{microtype} 
\usepackage{balance}
\usepackage{array}
\usepackage{threeparttable}
\usepackage[normalem]{ulem}

\input{include/vmr-symbols-vecbold}
\input{include/standard-macros}

\input{include/defs}

\safemath{\Tran}{\textnormal{T}}
\safemath{\Herm}{\textnormal{H}}
\usepackage{framed}

\IEEEoverridecommandlockouts
\allowdisplaybreaks 

\begin{document}

\bstctlcite{IEEEexample:BSTcontrol}

\title{A 1.2\,$\textnormal{mm}^\textnormal{2}$ 416\,mW 1.44\,M\,mat/s 64$\times$16 Matrix Preprocessing ASIC for Massive MIMO in 22FDX}

\author{\IEEEauthorblockN{Darja Nonaca and Christoph Studer}\\[-0.2cm]
\thanks{The work of DN and CS was supported in part by an ETH grant. Contact author: D.~Nonaca (e-mail: dnonaca@iis.ee.ethz.ch).}
\thanks{The authors would like to thank GlobalFoundries for providing silicon fabrication through the 22FDX University Program. The authors also thank Seyed Hadi Mirfarshbafan for his assistance with the ASIC design flow and Gian Marti for comments on this paper.}
\IEEEauthorblockA{\em Department of Information Technology and Electrical Engineering, ETH Zurich, Switzerland
} 
}

\maketitle

\input{0-abstract}
\input{1-introduction}

\input{2-algorithm}

\input{3-vlsi}

\input{4-implementation_results}

\input{5-conclusion}

 
\balance
\bibliographystyle{IEEEtran}
\bibliography{bib/confs-jrnls,bib/IEEEabrv,bib/publishers,bib/vipbib}
\balance

\end{document}

%% file: include/vmr-symbols-vecbold.tex
\usepackage{amssymb}
\usepackage{amsfonts}
\usepackage{mathrsfs}
\usepackage{xspace}
\usepackage{bm}
\usepackage{upgreek}

\newcommand{\safemath}[2]{\newcommand{#1}{\ensuremath{#2}\xspace}}

\safemath{\bma}{\mathbf{a}}
\safemath{\bmb}{\mathbf{b}}
\safemath{\bmc}{\mathbf{c}}
\safemath{\bmd}{\mathbf{d}}
\safemath{\bme}{\mathbf{e}}
\safemath{\bmf}{\mathbf{f}}
\safemath{\bmg}{\mathbf{g}}
\safemath{\bmh}{\mathbf{h}}
\safemath{\bmi}{\mathbf{i}}
\safemath{\bmj}{\mathbf{j}}
\safemath{\bmk}{\mathbf{k}}
\safemath{\bml}{\mathbf{l}}
\safemath{\bmm}{\mathbf{m}}
\safemath{\bmn}{\mathbf{n}}
\safemath{\bmo}{\mathbf{o}}
\safemath{\bmp}{\mathbf{p}}
\safemath{\bmq}{\mathbf{q}}
\safemath{\bmr}{\mathbf{r}}
\safemath{\bms}{\mathbf{s}}
\safemath{\bmt}{\mathbf{t}}
\safemath{\bmu}{\mathbf{u}}
\safemath{\bmv}{\mathbf{v}}
\safemath{\bmw}{\mathbf{w}}
\safemath{\bmx}{\mathbf{x}}
\safemath{\bmy}{\mathbf{y}}
\safemath{\bmz}{\mathbf{z}}
\safemath{\bmzero}{\mathbf{0}}
\safemath{\bmone}{\mathbf{1}}

\bmdefine{\biad}{a}
\bmdefine{\bibd}{b}
\bmdefine{\bicd}{c}
\bmdefine{\bidd}{d}
\bmdefine{\bied}{e}
\bmdefine{\bifd}{f}
\bmdefine{\bigd}{g}
\bmdefine{\bihd}{h}
\bmdefine{\biid}{i}
\bmdefine{\bijd}{j}
\bmdefine{\bikd}{k}
\bmdefine{\bild}{l}
\bmdefine{\bimd}{m}
\bmdefine{\bind}{n}
\bmdefine{\biod}{o}
\bmdefine{\bipd}{p}
\bmdefine{\biqd}{q}
\bmdefine{\bird}{r}
\bmdefine{\bisd}{s}
\bmdefine{\bitd}{t}
\bmdefine{\biud}{u}
\bmdefine{\bivd}{v}
\bmdefine{\biwd}{w}
\bmdefine{\bixd}{x}
\bmdefine{\biyd}{y}
\bmdefine{\bizd}{z}

\bmdefine{\bixid}{\xi}
\bmdefine{\bilambdad}{\lambda}
\bmdefine{\bimud}{\mu}
\bmdefine{\bithetad}{\theta}
\bmdefine{\biphid}{\phi}
\bmdefine{\bideltad}{\delta}

\safemath{\bmia}{\biad}
\safemath{\bmib}{\bibd}
\safemath{\bmic}{\bicd}
\safemath{\bmid}{\bidd}
\safemath{\bmie}{\bied}
\safemath{\bmif}{\bifd}
\safemath{\bmig}{\bigd}
\safemath{\bmih}{\bihd}
\safemath{\bmii}{\biid}
\safemath{\bmij}{\bijd}
\safemath{\bmik}{\bikd}
\safemath{\bmil}{\bild}
\safemath{\bmim}{\bimd}
\safemath{\bmin}{\bind}
\safemath{\bmio}{\biod}
\safemath{\bmip}{\bipd}
\safemath{\bmiq}{\biqd}
\safemath{\bmir}{\bird}
\safemath{\bmis}{\bisd}
\safemath{\bmit}{\bitd}
\safemath{\bmiu}{\biud}
\safemath{\bmiv}{\bivd}
\safemath{\bmiw}{\biwd}
\safemath{\bmix}{\bixd}
\safemath{\bmiy}{\biyd}
\safemath{\bmiz}{\bizd}

\safemath{\bmxi}{\bixid}
\safemath{\bmlambda}{\bilambdad}
\safemath{\bmmu}{\bimud}
\safemath{\bmtheta}{\bithetad}
\safemath{\bmphi}{\biphid}
\safemath{\bmdelta}{\bideltad}

\safemath{\bA}{\mathbf{A}}
\safemath{\bB}{\mathbf{B}}
\safemath{\bC}{\mathbf{C}}
\safemath{\bD}{\mathbf{D}}
\safemath{\bE}{\mathbf{E}}
\safemath{\bF}{\mathbf{F}}
\safemath{\bG}{\mathbf{G}}
\safemath{\bH}{\mathbf{H}}
\safemath{\bI}{\mathbf{I}}
\safemath{\bJ}{\mathbf{J}}
\safemath{\bK}{\mathbf{K}}
\safemath{\bL}{\mathbf{L}}
\safemath{\bM}{\mathbf{M}}
\safemath{\bN}{\mathbf{N}}
\safemath{\bO}{\mathbf{O}}
\safemath{\bP}{\mathbf{P}}
\safemath{\bQ}{\mathbf{Q}}
\safemath{\bR}{\mathbf{R}}
\safemath{\bS}{\mathbf{S}}
\safemath{\bT}{\mathbf{T}}
\safemath{\bU}{\mathbf{U}}
\safemath{\bV}{\mathbf{V}}
\safemath{\bW}{\mathbf{W}}
\safemath{\bX}{\mathbf{X}}
\safemath{\bY}{\mathbf{Y}}
\safemath{\bZ}{\mathbf{Z}}

\safemath{\bZero}{\mathbf{0}}
\safemath{\bOne}{\mathbf{1}}
\safemath{\bDelta}{\mathbf{\Delta}}
\safemath{\bLambda}{\mathbf{\UpLambda}}
\safemath{\bPhi}{\mathbf{\Upphi}}
\safemath{\bSigma}{\mathbf{\Upsigma}}
\safemath{\bOmega}{\mathbf{\Upomega}}
\safemath{\bTheta}{\mathbf{\Uptheta}}

\bmdefine{\biAd}{A}
\bmdefine{\biBd}{B}
\bmdefine{\biCd}{C}
\bmdefine{\biDd}{D}
\bmdefine{\biEd}{E}
\bmdefine{\biFd}{F}
\bmdefine{\biGd}{G}
\bmdefine{\biHd}{H}
\bmdefine{\biId}{I}
\bmdefine{\biJd}{J}
\bmdefine{\biKd}{K}
\bmdefine{\biLd}{L}
\bmdefine{\biMd}{M}
\bmdefine{\biOd}{N}
\bmdefine{\biPd}{O}
\bmdefine{\biQd}{P}
\bmdefine{\biRd}{R}
\bmdefine{\biSd}{S}
\bmdefine{\biTd}{T}
\bmdefine{\biUd}{U}
\bmdefine{\biVd}{V}
\bmdefine{\biWd}{W}
\bmdefine{\biXd}{X}
\bmdefine{\biYd}{Y}
\bmdefine{\biZd}{Z}

\bmdefine{\biDelta}{\Delta}
\bmdefine{\biLambda}{\Lambda}
\bmdefine{\biPhi}{\Phi}
\bmdefine{\biSigma}{\Sigma}
\bmdefine{\biOmega}{\Omega}
\bmdefine{\biTheta}{\Theta}

\safemath{\bimA}{\biAd}
\safemath{\bimB}{\biBd}
\safemath{\bimC}{\biCd}
\safemath{\bimD}{\biDd}
\safemath{\bimE}{\biEd}
\safemath{\bimF}{\biFd}
\safemath{\bimG}{\biGd}
\safemath{\bimH}{\biHd}
\safemath{\bimI}{\biId}
\safemath{\bimJ}{\biJd}
\safemath{\bimK}{\biKd}
\safemath{\bimL}{\biLd}
\safemath{\bimM}{\biMd}
\safemath{\bimN}{\biNd}
\safemath{\bimO}{\biOd}
\safemath{\bimP}{\biPd}
\safemath{\bimQ}{\biQd}
\safemath{\bimR}{\biRd}
\safemath{\bimS}{\biSd}
\safemath{\bimT}{\biTd}
\safemath{\bimU}{\biUd}
\safemath{\bimV}{\biVd}
\safemath{\bimW}{\biWd}
\safemath{\bimX}{\biXd}
\safemath{\bimY}{\biYd}
\safemath{\bimZ}{\biZd}

\safemath{\bimDelta}{\biDelta}
\safemath{\bimLambda}{\biLambda}
\safemath{\bimPhi}{\biPhi}
\safemath{\bimSigma}{\biSigma}
\safemath{\bimOmega}{\biOmega}
\safemath{\bimTheta}{\biTheta}

\safemath{\setA}{\mathcal{A}}
\safemath{\setB}{\mathcal{B}}
\safemath{\setC}{\mathcal{C}}
\safemath{\setD}{\mathcal{D}}
\safemath{\setE}{\mathcal{E}}
\safemath{\setF}{\mathcal{F}}
\safemath{\setG}{\mathcal{G}}
\safemath{\setH}{\mathcal{H}}
\safemath{\setI}{\mathcal{I}}
\safemath{\setJ}{\mathcal{J}}
\safemath{\setK}{\mathcal{K}}
\safemath{\setL}{\mathcal{L}}
\safemath{\setM}{\mathcal{M}}
\safemath{\setN}{\mathcal{N}}
\safemath{\setO}{\mathcal{O}}
\safemath{\setP}{\mathcal{P}}
\safemath{\setQ}{\mathcal{Q}}
\safemath{\setR}{\mathcal{R}}
\safemath{\setS}{\mathcal{S}}
\safemath{\setT}{\mathcal{T}}
\safemath{\setU}{\mathcal{U}}
\safemath{\setV}{\mathcal{V}}
\safemath{\setW}{\mathcal{W}}
\safemath{\setX}{\mathcal{X}}
\safemath{\setY}{\mathcal{Y}}
\safemath{\setZ}{\mathcal{Z}}
\safemath{\emptySet}{\varnothing}

\safemath{\colA}{\mathscr{A}}
\safemath{\colB}{\mathscr{B}}
\safemath{\colC}{\mathscr{C}}
\safemath{\colD}{\mathscr{D}}
\safemath{\colE}{\mathscr{E}}
\safemath{\colF}{\mathscr{F}}
\safemath{\colG}{\mathscr{G}}
\safemath{\colH}{\mathscr{H}}
\safemath{\colI}{\mathscr{I}}
\safemath{\colJ}{\mathscr{J}}
\safemath{\colK}{\mathscr{K}}
\safemath{\colL}{\mathscr{L}}
\safemath{\colM}{\mathscr{M}}
\safemath{\colN}{\mathscr{N}}
\safemath{\colO}{\mathscr{O}}
\safemath{\colP}{\mathscr{P}}
\safemath{\colQ}{\mathscr{Q}}
\safemath{\colR}{\mathscr{R}}
\safemath{\colS}{\mathscr{S}}
\safemath{\colT}{\mathscr{T}}
\safemath{\colU}{\mathscr{U}}
\safemath{\colV}{\mathscr{V}}
\safemath{\colW}{\mathscr{W}}
\safemath{\colX}{\mathscr{X}}
\safemath{\colY}{\mathscr{Y}}
\safemath{\colZ}{\mathscr{Z}}

\safemath{\opA}{\mathbb{A}}
\safemath{\opB}{\mathbb{B}}
\safemath{\opC}{\mathbb{C}}
\safemath{\opD}{\mathbb{D}}
\safemath{\opE}{\mathbb{E}}
\safemath{\opF}{\mathbb{F}}
\safemath{\opG}{\mathbb{G}}
\safemath{\opH}{\mathbb{H}}
\safemath{\opI}{\mathbb{I}}
\safemath{\opJ}{\mathbb{J}}
\safemath{\opK}{\mathbb{K}}
\safemath{\opL}{\mathbb{L}}
\safemath{\opM}{\mathbb{M}}
\safemath{\opN}{\mathbb{N}}
\safemath{\opO}{\mathbb{O}}
\safemath{\opP}{\mathbb{P}}
\safemath{\opQ}{\mathbb{Q}}
\safemath{\opR}{\mathbb{R}}
\safemath{\opS}{\mathbb{S}}
\safemath{\opT}{\mathbb{T}}
\safemath{\opU}{\mathbb{U}}
\safemath{\opV}{\mathbb{V}}
\safemath{\opW}{\mathbb{W}}
\safemath{\opX}{\mathbb{X}}
\safemath{\opY}{\mathbb{Y}}
\safemath{\opZ}{\mathbb{Z}}
\safemath{\opZero}{\mathbb{O}}
\safemath{\identityop}{\opI}

\safemath{\veca}{\bma}
\safemath{\vecb}{\bmb}
\safemath{\vecc}{\bmc}
\safemath{\vecd}{\bmd}
\safemath{\vece}{\bme}
\safemath{\vecf}{\bmf}
\safemath{\vecg}{\bmg}
\safemath{\vech}{\bmh}
\safemath{\veci}{\bmi}
\safemath{\vecj}{\bmj}
\safemath{\veck}{\bmk}
\safemath{\vecl}{\bml}
\safemath{\vecm}{\bmm}
\safemath{\vecn}{\bmn}
\safemath{\veco}{\bmo}
\safemath{\vecp}{\bmp}
\safemath{\vecq}{\bmq}
\safemath{\vecr}{\bmr}
\safemath{\vecs}{\bms}
\safemath{\vect}{\bmt}
\safemath{\vecu}{\bmu}
\safemath{\vecv}{\bmv}
\safemath{\vecw}{\bmw}
\safemath{\vecx}{\bmx}
\safemath{\vecy}{\bmy}
\safemath{\vecz}{\bmz}

\safemath{\veczero}{\bmzero}
\safemath{\vecone}{\bmone}
\safemath{\vecxi}{\bmxi}
\safemath{\veclambda}{\bmlambda}
\safemath{\vecmu}{\bmmu}
\safemath{\vectheta}{\bmtheta}
\safemath{\vecphi}{\bmphi}
\safemath{\vecdelta}{\bmdelta}

\safemath{\matA}{\bA}
\safemath{\matB}{\bB}
\safemath{\matC}{\bC}
\safemath{\matD}{\bD}
\safemath{\matE}{\bE}
\safemath{\matF}{\bF}
\safemath{\matG}{\bG}
\safemath{\matH}{\bH}
\safemath{\matI}{\bI}
\safemath{\matJ}{\bJ}
\safemath{\matK}{\bK}
\safemath{\matL}{\bL}
\safemath{\matM}{\bM}
\safemath{\matN}{\bN}
\safemath{\matO}{\bO}
\safemath{\matP}{\bP}
\safemath{\matQ}{\bQ}
\safemath{\matR}{\bR}
\safemath{\matS}{\bS}
\safemath{\matT}{\bT}
\safemath{\matU}{\bU}
\safemath{\matV}{\bV}
\safemath{\matW}{\bW}
\safemath{\matX}{\bX}
\safemath{\matY}{\bY}
\safemath{\matZ}{\bZ}
\safemath{\matzero}{\bmzero}

\safemath{\matDelta}{\bDelta}
\safemath{\matLambda}{\bLambda}
\safemath{\matPhi}{\bPhi}
\safemath{\matSigma}{\bSigma}
\safemath{\matOmega}{\bOmega}
\safemath{\matTheta}{\bTheta}

\safemath{\matidentity}{\matI}
\safemath{\matone}{\matO}

\safemath{\rnda}{A}
\safemath{\rndb}{B}
\safemath{\rndc}{C}
\safemath{\rndd}{D}
\safemath{\rnde}{E}
\safemath{\rndf}{F}
\safemath{\rndg}{G}
\safemath{\rndh}{H}
\safemath{\rndi}{I}
\safemath{\rndj}{J}
\safemath{\rndk}{K}
\safemath{\rndl}{L}
\safemath{\rndm}{M}
\safemath{\rndn}{N}
\safemath{\rndo}{O}
\safemath{\rndp}{P}
\safemath{\rndq}{Q}
\safemath{\rndr}{R}
\safemath{\rnds}{S}
\safemath{\rndt}{T}
\safemath{\rndu}{U}
\safemath{\rndv}{V}
\safemath{\rndw}{W}
\safemath{\rndx}{X}
\safemath{\rndy}{Y}
\safemath{\rndz}{Z}

\safemath{\rveca}{\bimA}
\safemath{\rvecb}{\bimB}
\safemath{\rvecc}{\bimC}
\safemath{\rvecd}{\bimD}
\safemath{\rvece}{\bimE}
\safemath{\rvecf}{\bimF}
\safemath{\rvecg}{\bimG}
\safemath{\rvech}{\bimH}
\safemath{\rveci}{\bimI}
\safemath{\rvecj}{\bimJ}
\safemath{\rveck}{\bimK}
\safemath{\rvecl}{\bimL}
\safemath{\rvecm}{\bimM}
\safemath{\rvecn}{\bimN}
\safemath{\rveco}{\bomO}
\safemath{\rvecp}{\bimP}
\safemath{\rvecq}{\bimQ}
\safemath{\rvecr}{\bimR}
\safemath{\rvecs}{\bimS}
\safemath{\rvect}{\bimT}
\safemath{\rvecu}{\bimU}
\safemath{\rvecv}{\bimV}
\safemath{\rvecw}{\bimW}
\safemath{\rvecx}{\bimX}
\safemath{\rvecy}{\bimY}
\safemath{\rvecz}{\bimZ}

\safemath{\rvecxi}{\bmxi}
\safemath{\rveclambda}{\bmlambda}
\safemath{\rvecmu}{\bmmu}
\safemath{\rvectheta}{\bmtheta}
\safemath{\rvecphi}{\bmphi}

\safemath{\rmatA}{\bimA}
\safemath{\rmatB}{\bimB}
\safemath{\rmatC}{\bimC}
\safemath{\rmatD}{\bimD}
\safemath{\rmatE}{\bimE}
\safemath{\rmatF}{\bimF}
\safemath{\rmatG}{\bimG}
\safemath{\rmatH}{\bimH}
\safemath{\rmatI}{\bimI}
\safemath{\rmatJ}{\bimJ}
\safemath{\rmatK}{\bimK}
\safemath{\rmatL}{\bimL}
\safemath{\rmatM}{\bimM}
\safemath{\rmatN}{\bimN}
\safemath{\rmatO}{\bimO}
\safemath{\rmatP}{\bimP}
\safemath{\rmatQ}{\bimQ}
\safemath{\rmatR}{\bimR}
\safemath{\rmatS}{\bimS}
\safemath{\rmatT}{\bimT}
\safemath{\rmatU}{\bimU}
\safemath{\rmatV}{\bimV}
\safemath{\rmatW}{\bimW}
\safemath{\rmatX}{\bimX}
\safemath{\rmatY}{\bimY}
\safemath{\rmatZ}{\bimZ}

\safemath{\rmatDelta}{\bimDelta}
\safemath{\rmatLambda}{\bimLambda}
\safemath{\rmatPhi}{\bimPhi}
\safemath{\rmatSigma}{\bimSigma}
\safemath{\rmatOmega}{\bimOmega}
\safemath{\rmatTheta}{\bimTheta}

%% file: include/standard-macros.tex
\usepackage{amssymb}
\usepackage{amsfonts}
\usepackage{mathrsfs}
\usepackage{xspace}
\usepackage{bm}
\usepackage{fancyref}
\usepackage{textcomp}

\usepackage{multirow}
\usepackage{stmaryrd}

\newenvironment{textbmatrix}{	\setlength{\arraycolsep}{2.5pt}%
								\big[\begin{matrix}}{\end{matrix}\big]%
								\raisebox{0.08ex}{\vphantom{M}}}

\def\be{\begin{equation}}
\def\ee{\end{equation}}
\def\een{\nonumber \end{equation}}
\def\mat{\begin{bmatrix}}
\def\emat{\end{bmatrix}}
\def\btm{\begin{textbmatrix}}
\def\etm{\end{textbmatrix}}

\def\ba#1\ea{\begin{align}#1\end{align}}
\def\bas#1\eas{\begin{align*}#1\end{align*}}
\def\bs#1\es{\begin{split}#1\end{split}}
\def\bg#1\eg{\begin{gather}#1\end{gather}}
\def\bml#1\eml{\begin{multline}#1\end{multline}}
\def\bi#1\ei{\begin{itemize}#1\end{itemize}}

\safemath{\dirac}{\delta}					
\safemath{\krond}{\dirac}					

\safemath{\upto}{\uparrow}
\safemath{\downto}{\downarrow}
\safemath{\iu}{j}							
\safemath{\ev}{\lambda}						
\safemath{\hilseqspace}{l^{2}}				
\newcommand{\banachfunspace}[1]{\setL^{#1}}	
\safemath{\hilfunspace}{\banachfunspace{2}}	

\safemath{\SNR}{\textit{SNR}} 				
\safemath{\PAR}{\textit{PAR}} 				
\safemath{\No}{N_0}							
\safemath{\Es}{E_s}							
\safemath{\Eb}{E_b}							
\safemath{\EbNo}{\frac{\Eb}{\No}}
\safemath{\EsNo}{\frac{\Es}{\No}}

\DeclareMathOperator{\CHop}{\ensuremath{\opH}} 
\safemath{\tvir}{\rndh_{\CHop}}				
\safemath{\tvtf}{\rndl_{\CHop}}				
\safemath{\spf}{\rnds_{\CHop}}				
\safemath{\bff}{H_{\CHop}}					

\safemath{\ircf}{r_{h}}						
\safemath{\tftvcf}{r_{s}}					
\safemath{\tfcf}{r_{l}}						
\safemath{\bfcf}{r_{H}}						

\safemath{\tcorr}{c_h}						
\safemath{\scf}{c_{s}}						
\safemath{\tfcorr}{c_{l}}					
\safemath{\fcorr}{c_{H}}						

\safemath{\mi}{I}							
\safemath{\capacity}{C}						

\safemath{\normal}{\mathcal{N}}			
\safemath{\jpg}{\mathcal{CN}}			
\safemath{\mchain}{\leftrightarrow}		

\safemath{\dB}{\,\mathrm{dB}}
\safemath{\dBm}{\,\mathrm{dBm}}
\safemath{\Hz}{\,\mathrm{Hz}}
\safemath{\kHz}{\,\mathrm{kHz}}
\safemath{\MHz}{\,\mathrm{MHz}}
\safemath{\GHz}{\,\mathrm{GHz}}
\safemath{\s}{\,\mathrm{s}}
\safemath{\ms}{\,\mathrm{ms}}
\safemath{\mus}{\,\mathrm{\text{\textmu}s}}
\safemath{\ns}{\,\mathrm{ns}}
\safemath{\ps}{\,\mathrm{ps}}
\safemath{\meter}{\,\mathrm{m}}
\safemath{\mm}{\,\mathrm{mm}}
\safemath{\cm}{\,\mathrm{cm}}
\safemath{\m}{\,\mathrm{m}}
\safemath{\W}{\,\mathrm{W}}
\safemath{\mW}{\, \mathrm{mW}}
\safemath{\J}{\,\mathrm{J}}
\safemath{\K}{\,\mathrm{K}}
\safemath{\bit}{\,\mathrm{bit}}
\safemath{\nat}{\,\mathrm{nat}}

\safemath{\define}{\triangleq}			

\safemath{\equivalent}{\sim}
\safemath{\distas}{\sim}					
\safemath{\sdiff}{\Delta}				

\safemath{\reals}{\mathbb{R}}
\safemath{\positivereals}{\reals_{+}}
\safemath{\integers}{\mathbb{Z}}
\safemath{\posint}{\integers_{+}}
\safemath{\naturals}{\mathbb{N}}
\safemath{\posnaturals}{\naturals_{+}}
\safemath{\complexset}{\mathbb{C}}
\safemath{\rationals}{\mathbb{Q}}

\newcommand*{\fancyrefapplabelprefix}{app}		
\newcommand*{\fancyrefthmlabelprefix}{thm}		
\newcommand*{\fancyreflemlabelprefix}{lem}		
\newcommand*{\fancyrefcorlabelprefix}{cor}		
\newcommand*{\fancyrefdeflabelprefix}{def}		
\newcommand*{\fancyrefproplabelprefix}{prop}		
\newcommand*{\fancyrefexmpllabelprefix}{exmpl}
\newcommand*{\fancyrefalglabelprefix}{alg}		
\newcommand*{\fancyreftbllabelprefix}{tbl}		

\frefformat{vario}{\fancyrefseclabelprefix}{Sec.~#1}
\frefformat{vario}{\fancyrefthmlabelprefix}{Theorem~#1}
\frefformat{vario}{\fancyreftbllabelprefix}{Table~#1}
\frefformat{vario}{\fancyreflemlabelprefix}{Lemma~#1}
\frefformat{vario}{\fancyrefcorlabelprefix}{Corollary~#1}
\frefformat{vario}{\fancyrefdeflabelprefix}{Definition~#1}
\frefformat{vario}{\fancyreffiglabelprefix}{Fig.~#1}
\frefformat{vario}{\fancyrefapplabelprefix}{Appendix~#1}
\frefformat{vario}{\fancyrefeqlabelprefix}{(#1)}
\frefformat{vario}{\fancyrefproplabelprefix}{Proposition~#1}
\frefformat{vario}{\fancyrefexmpllabelprefix}{Example~#1}
\frefformat{vario}{\fancyrefalglabelprefix}{Alg.~#1}

%% file: include/defs.tex
\safemath{\dictab}{[\,\dicta\,\,\dictb\,]}

\safemath{\ysig}{\bmy}
\safemath{\ysighat}{\hat{\ysig}}
\safemath{\ysigdim}{M}
\safemath{\xsig}{\bmx}
\safemath{\xsigdim}{N}
\safemath{\nx}{n_x}
\safemath{\zsig}{\bmz}
\safemath{\zsigdim}{\ysigdim}
\safemath{\rsig}{\bmr}
\safemath{\Adict}{\bA}
\safemath{\Adicttilde}{\widetilde{\Adict}}
\safemath{\Adictdim}{\outputdim\times\xsigdim}
\safemath{\avec}{\bma}
\safemath{\avectilde}{\tilde{\avec}}
\safemath{\Bdict}{\bB}
\safemath{\Bdicttilde}{\widetilde{\Bdict}}
\safemath{\Cdict}{\bC}
\safemath{\cvec}{\bmc}
\safemath{\Ddict}{\bD}
\safemath{\Ddictdim}{\ysigdim\times\xsigdim}
\safemath{\dvec}{\bmd}
\safemath{\Ddicttilde}{\widetilde{\bD}}
\safemath{\Bonb}{\bB}
\safemath{\bvec}{\bmb}
\safemath{\Bonbdim}{\ysigdim\times\ysigdim}
\safemath{\noise}{\bmn}
\safemath{\noisedim}{\ysigim}
\safemath{\err}{\bme}
\safemath{\errdim}{\ysigdim}
\safemath{\errset}{\setE}
\safemath{\nerr}{n_e}
\safemath{\delop}{\bP_\errset}
\safemath{\delopc}{\bP_{{\errset}^c}}

\safemath{\cplxi}{\imath}
\safemath{\cplxj}{\jmath}

\safemath{\dict}{\matD}
\safemath{\inputdim}{N}		
\safemath{\outputdim}{M}		
\safemath{\sparsity}{S}	
\safemath{\inputdimA}{{N_a}}	
\safemath{\inputdimB}{{N_b}}	
\safemath{\elemA}{{n_a}}	
\safemath{\elemB}{{n_b}}	
\safemath{\resA}{\matR_a}	
\safemath{\resB}{\matR_b}	
\safemath{\subD}{\matS} 
\safemath{\subA}{\matS_a} 
\safemath{\subB}{\matS_b} 
\safemath{\dicta}{\matA} 	
\safemath{\dictb}{\matB} 	
\safemath{\hollowS}{H}
\safemath{\hollowA}{H_a}
\safemath{\hollowB}{H_b}
\safemath{\cross}{Z}
\safemath{\coh}{\mu_d}			
\safemath{\coha}{\mu_a}			
\safemath{\cohb}{\mu_b}			
\safemath{\mubs}{\nu}	
\safemath{\cohm}{\mu_m} 
\safemath{\dictset}{\setD}	
\safemath{\dictsetp}{\dictset(\coh,\coha,\cohb)}	
\safemath{\dictsetgen}{\dictset_\text{gen}}
\safemath{\dictsetgenp}{\dictsetgen(\coh)}
\safemath{\dictsetonb}{\dictset_\text{onb}}
\safemath{\dictsetonbp}{\dictsetonb(\coh)}

\safemath{\leftside}{U}
\safemath{\rightsideA}{R_a}
\safemath{\rightsideB}{R_b}

\safemath{\indexS}{\setI_S} 

\safemath{\na}{n_a}			
\safemath{\nb}{n_b}			
\safemath{\coeffa}{p_i}	
\safemath{\coeffb}{q_j}	
\safemath{\seta}{\setP}		
\safemath{\setb}{\setQ}     
\safemath{\setw}{\setW}	
\safemath{\setz}{\setZ}	
\safemath{\cola}{\veca}		
\safemath{\colb}{\vecb}		
\safemath{\cold}{\vecd}		
\safemath{\inputvec}{\vecx} 	
\safemath{\error}{\vece}	
\safemath{\noiseout}{\vecz} 	
\safemath{\inputvecel}{x}
\safemath{\inputveca}{\vecx_a}
\safemath{\inputvecb}{\vecx_b}
\safemath{\outputvec}{\vecy}	
\safemath{\lambdamin}{\lambda_{\mathrm{min}}}

\safemath{\elltwo}{\ell_2}
\safemath{\ellone}{\ell_1}
\safemath{\ellzero}{\ell_0}
\safemath{\ellinf}{\ell_\infty}
\safemath{\ellinftilde}{\ell_{\widetilde\infty}}
\safemath{\licard}{Z(\coh,\coha,\cohb)}
\safemath{\xsol}{\hat{x}}
\safemath{\xbord}{x_b}		
\safemath{\xstat}{x_s}		
\safemath{\xstatLone}{\tilde{x}_s}
\safemath{\order}{\mathcal{O}} 
\safemath{\scales}{\Theta} 
\safemath{\ones}{\mathbf{1}} 
\safemath{\zeroes}{\mathbf{0}} 
\safemath{\thlone}{\kappa(\coh,\cohb)} 
\safemath{\constoneA}{\delta} 
\safemath{\constoneB}{\epsilon} 
\safemath{\nlarge}{L}				   
\safemath{\sumlarge}{S_\nlarge}
\safemath{\maxlarger}{P_\nlarge}	   
\safemath{\Pzero}{\textrm{P0}}	
\safemath{\Pone}{\textrm{P1}}
\safemath{\vecfir}{\vecw}			 
\safemath{\vecsec}{\vecz}
\safemath{\elvecfir}{w}              
\safemath{\elvecsec}{z}				 
\safemath{\nlargefir}{n}
\safemath{\normout}{\gamma}
\safemath{\auxfun}{h}
\safemath{\supp}{\textrm{supp}}

\safemath{\indexa}{\ell}
\safemath{\indexb}{r}
\safemath{\indexc}{i}
\safemath{\indexd}{j}

\safemath{\project}{P}

%% file: 0-abstract.tex

\begin{abstract}
Massive multiuser (MU) multiple-input multiple-output (MIMO) enables concurrent transmission of multiple users to a multi-antenna basestation (BS). To detect the users' data using linear equalization, the BS must perform preprocessing, which requires, among other tasks, the inversion of a matrix whose dimension equals the number of user data streams. Explicit inversion of large matrices is notoriously difficult to implement due to high complexity, stringent data dependencies that lead to high latency, and high numerical precision requirements. 
We propose a novel preprocessing architecture based on the block-LDL matrix factorization, which improves parallelism and, hence, reduces latency. 
We demonstrate the effectiveness of our architecture through (i) massive MU-MIMO system simulations with mmWave channel vectors and (ii) measurements of a 22FDX ASIC, which is, to our knowledge, the first fabricated preprocessing engine for massive MU-MIMO with 64 BS antennas and 16 single-antenna users. Our ASIC reaches a clock frequency of 870\,MHz while consuming 416\,mW. 
At its peak throughput, the ASIC preprocesses 1.44\,M $\boldsymbol{64\times 16}$ matrices per second at a latency of only 0.7\,\textmu{}s.
\end{abstract}

%% file: 1-introduction.tex

\section{Introduction}

Modern wireless communication systems leverage massive multiple-input multiple-output (MIMO) to enable multiuser (MU) communication at high data rates~\cite{busari}. 
To enable efficient hardware implementation of data detection at the basestation (BS), one typically resorts to linear methods, such as linear minimum mean square error (LMMSE)-based equalization~\cite{juntti}. 
The complexity of such approaches, however, grows quickly for systems that must support a large number of simultaneously-transmitting users. 
In particular, the complexity of preprocessing, which computes the LMMSE filter matrix every time the channel changes, grows cubically in the number of users for all methods that have been implemented in hardware. 
Besides minimizing complexity, the preprocessing latency must be kept at a minimum to adhere to the stringent latency constraints of modern wireless systems. 
While approximate preprocessing methods that scale only quadratically in the number of users have been proposed~\cite{mwu}, they only perform well (i) if the number of BS antennas is substantially larger than the number of users and (ii) the users' channels are sufficiently distinct.  
Therefore, efficient and also exact matrix preprocessing algorithms and hardware implementations are crucial to meeting the latency and quality constraints of massive MU-MIMO systems.

\subsection{Contributions}
We propose the first fabricated ASIC of an exact matrix preprocessing engine for 
LMMSE-based data detection in massive MU-MIMO systems with 64 BS antennas and 16 single-antenna users. Our architecture carries out the following steps: (i) Gram-matrix computation, (ii) block-LDL (BLDL) matrix factorization, and (iii) backward substitution. To reduce complexity, we utilize the method from~\cite{clasen} to skip the otherwise necessary forward-substitution step. 
In contrast to other matrix-factorization methods, our BLDL-based architecture processes more data items in parallel, which reduces latency. 
To reduce silicon area, steps (i) and (ii) share the same hardware resources. 
A comparison with existing designs reveals that 
our fabricated and measured ASIC outperforms other
designs in terms of throughput, area, latency, and/or error-rate performance.

\subsection{Relevant Prior Work}

A variety of algorithms for explicit matrix inversion exist, such as methods based on  the Cholesky, LU, LDL, and QR matrix factorizations~\cite{golub}.
Several hardware architectures for preprocessing and LMMSE-based data detection in small-scale MIMO systems have been proposed: 
Reference~\cite{burg06} implements matrix inversion using rank-1 updates; references~\cite{studer} and~\cite{auras} perform an LU and LDL factorization, respectively, followed by forward and backward substitution; 
and references~\cite{omran,singh} perform a QR factorization followed by inversion of the triangular matrix.
For massive MU-MIMO systems, reference~\cite{mahdavi} provides synthesis results of a $128\times 16$ preprocessing engine that uses the Cholesky decomposition followed by forward and backward substitution.
References~\cite{abbas, prabhu} implement an approximate matrix inversion based on the  Neumann series, which reduces complexity but sacrifices error-rate performance.
In contrast, we propose an \emph{exact} $64\times16$ BLDL-based matrix-preprocessing engine that avoids backward substitution, which reduces latency and complexity. Furthermore, we provide measurement results of a fabricated~ASIC in 22FDX. 

\subsection{Notation}
Boldface lowercase and uppercase letters represent column vectors and matrices, respectively. 
For a matrix $\bG$ partitioned into $2\times2$ blocks, $\bG_{ij}\in\complexset^{2\times2}$ is the submatrix formed by the elements of $\bG$ consisting of the rows $(2(i-1)+1:2(i-1)+2)$ and columns $(2(j-1)+1:2(j-1)+2)$.
The Hermitian transpose of $\bG$ is $\bG^\Herm$, and the entry on the $m$th row and $n$th column is $g_{mn}$. Complex conjugation is indicated by the superscript $^*$.
The $N\times N$ identity matrix is $\bI_N$.

%% file: 2-algorithm.tex
\section{System Model and BLDL-based Preprocessing}
\label{sec:algo} 

\subsection{System Model and LMMSE-based Data Detection}
We focus on the massive MU-MIMO uplink, in which $U$ single-antenna UEs transmit data to a $B$-antenna BS. 
We model the frequency-flat input-output relation as $\bmy = \bH\bms + \bmn$,
where $\bmy\in \mathbb{C}^B$ is the received vector at the BS, $\bH\in \mathbb{C}^{B\times U}$ is the channel matrix, $\bms\in\setX^U$ is the transmit symbol vector with entries taken from a constellation $\setX$ whose energy is normalized to $\Es$, and $\bmn\in \mathbb{C}^B$ is i.i.d.\ circularly-symmetric complex Gaussian noise with variance $\No$ per entry. 

Data detection deals with recovering the transmit vector~$\bms$ from $\bmy$ and (an estimate of) $\bH$.
LMMSE-based methods perform data detection in two phases:
(i) \emph{preprocessing} first calculates and then inverts the matrix
\begin{align} \label{eq:regularizedgrammatrix}
    \bA = \textstyle \bH^\Herm\bH + \frac{\No}{\Es}\bI_U,
\end{align}
whenever the channel matrix $\bH$ changes; and (ii) \emph{equalization} is  carried out for every transmit vector $\bms$ according to $\hat{\bms}_\textnormal{LMMSE} = \bA^{-1}\bH^\Herm\bmy$. 
In what follows, we focus on the preprocessing phase as it dominates complexity and latency.


\begin{algorithm}[tp]
\caption{Block-LDL (BLDL) factorization~\cite{stan}}\label{alg:blockchol}
\begin{algorithmic}
\STATE \textbf{input:} $\bA\in \mathbb{C}^{U\times U}$ partitioned into $2\times2$ blocks; $N = U/2$ \\
\FOR{$j=1$ to $N$} 
    \STATE $\bD_{jj} = \bA_{jj} - \sum_{k=1}^{j-1} \bL_{jk}\bD_{kk}\bL_{jk}^{\Herm}$ \
    \FOR{$i=j+1$ to $N$} 
        \STATE $\bL_{ij} = (\bA_{ij} - \sum_{k=1}^{j-1}  \bL_{ik}\bD_{kk}\bL_{jk}^{\Herm}) \bD_{jj}^{-1}$ \
    \ENDFOR
\ENDFOR
\STATE \textbf{output: $\bL, \bD^{-1}$} 
\end{algorithmic}
\end{algorithm}

\subsection{BLDL-based Matrix Preprocessing}
\label{sec:inversionprocedure}

After computing $\bA$ as in \fref{eq:regularizedgrammatrix}, we factorize $\bA=\bL\bD\bL^{\Herm}$, where $\bL$ is a lower-triangular with $2\times2$ identity matrices on the diagonal and $\bD$ is a block diagonal matrix also consisting of $2\times2$ blocks.
To improve parallelism, we utilize the BLDL factorization from~\cite{stan}, which is summarized in Alg.~\ref{alg:blockchol}. 
In our architecture, we partition $\bA$ into $2\times2$ submatrices and the $2\times2$ submatrix inversions $\bD_{jj}^{-1},\,j=1,\ldots,\frac{U}{2}$ in Alg.~\ref{alg:blockchol} are calculated efficiently via direct inversion~\cite{golub}
\begin{align}\label{eq:3}
\bD_{jj}^{-1} = \left[
\begin{array}{cc} 
a   & b  \\
b^* & d   \\
\end{array}
\right] ^{-1}= 
\frac{1}{\Delta}\left[ 
\begin{array}{cc}
 d   & -b  \\
-b^* &  a  \\
\end{array}
\right]\!,
\end{align}
where $\Delta=ad-bb^*$ is the determinant of $\bD_{jj}$.
After the BLDL factorization, we can rewrite $\bA^{-1}$ as
\begin{align} \label{eq:e1}
    \bA^{-1}  = (\bL\bD\bL^\Herm)^{-1} =(\bL^\Herm)^{-1} \bD^{-1}\bL^{-1}.
\end{align}
\noindent
By multiplying \fref{eq:e1} from the left by $\bL^\Herm$, we now solve 
\begin{align} \label{eq:e4}
    \bL^\Herm\bX = \bD^{-1}\bL^{-1}
\end{align}
for $\bX$, where the solution $\hat\bX = \bA^{-1}$ will be the desired inverse.
To reduce complexity, we follow the idea of~\cite{clasen} to solve for~$\bX$ without computing the inverse $\bL^{-1}$. 
To illustrate this idea, consider the following simplified $3\times3$ example of  \fref{eq:e4}
\begin{align}\label{eq:e5}
\hspace{-0.2cm}
\resizebox{0.94\columnwidth}{!}{$\left[
\hspace{-4pt}
\begin{array}{ccc}
1 & l_{21}^{*}   & l_{31}^{*} \\
0 & 1            & l_{32}^{*} \\
0 & 0            & 1\\
\end{array}
\hspace{-4pt}
\right]
\hspace{-5pt}
\left[
\hspace{-4pt}
\begin{array}{ccc}
x_{11}     & x_{12}     & x_{13}\\
x_{12}^{*} & x_{22}     & x_{23}\\
x_{13}^{*} & x_{23}^{*} & x_{33}\\
\end{array}
\hspace{-4pt}
\right]
\hspace{-4pt}
 = 
\hspace{-4pt}
\left[
\hspace{-4pt}
\begin{array}{ccc}
 d_{1}^{-1}          & 0                   & 0\\
 \alpha d_{2}^{-1}   & d_{2}^{-1}          & 0\\
 \beta  d_{3}^{-1}   & \gamma d_{3}^{-1}   & d_{3}^{-1}\\
\end{array}
\hspace{-4pt}
\right]\!\!,
$}
\end{align}
where (i) we use the fact that the inverse of a lower-triangular matrix is lower triangular and (ii) $\alpha$, $\beta$, and $\gamma$ are the off-diagonal elements of $\bL^{-1}$.
The process starts by solving for the third column of $\bX$ from bottom to top using backward substitution. This is equivalent to solving the following equations one after the other: \\[-0.4cm]
\begin{align}\label{eq:e7}
\begin{array}{rcc}
x_{33} & = & d_{3}^{-1} \\
x_{23} + l_{32}^{*}x_{33} & =& 0\phantom{.} \\
x_{13} + l_{21}^{*}x_{23} + l_{31}^{*}x_{33} & = & 0.
\end{array}
\end{align}
Once $x_{33}$, $x_{23}$, and $x_{13} $ have been computed, one proceeds analogously by computing the second column of $\bX$ by solving
\begin{align}\label{eq:e8}
\begin{array}{rcc}
x_{22} + l_{32}^{*}x_{23}^{*} & = & d_{2}^{-1} \\
x_{12} + l_{21}^{*}x_{22} + l_{31}^{*}x_{23}^{*} & = & 0. \\
\end{array}
\end{align}
\noindent
Finally, one proceeds with the first column to solve for $x_{11}$: 
\begin{align}\label{eq:e9}
\begin{array}{rcc}
x_{11} + l_{21}^{*}x_{12}^{*} + l_{31}^{*}x_{13}^{*} & = &  d_{1}^{-1}.
\end{array}
\end{align}
We reiterate that this process for solving for $\bX=\bA^{-1}$ avoids inverting $\bL$ as the off-diagonal elements $\alpha$, $\beta$, and $\gamma$ are unused.

%% file: 3-vlsi.tex

\section{VLSI Architecture}
\label{sec:vlsi}

\subsection{Architecture Overview}
\begin{figure}[tp]
\centering
\includegraphics[width=0.99\columnwidth]{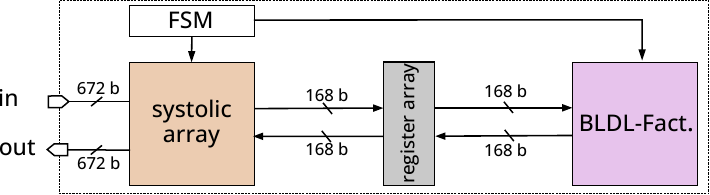}
\vspace{-0.6cm}
\caption{Top-level architecture of the implemented preprocessing engine. The bus width of the input and output data accommodates the size of a row of the channel matrix~$\bH$ (16 complex values of 21 bits per part). The bus at the interface with the register array fits a $2\times2$ matrix with $4$ complex values.}
\label{fig:top_level}
\end{figure} 

\begin{figure}[tp]
\centering
\includegraphics[width=1\columnwidth]{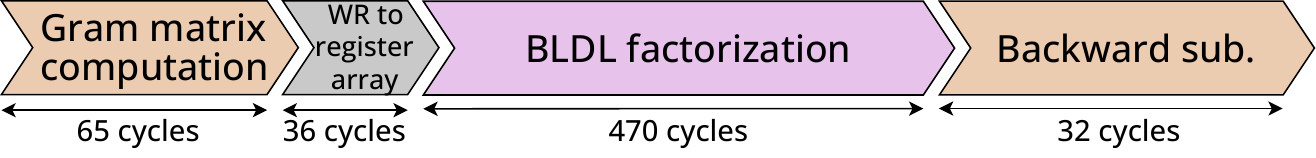}
\vspace{-0.7cm}
\caption{Schedule (in clock cycles) of each preprocessing step.}
\label{fig:mothra_schedule}
\end{figure} 

Fig.~\ref{fig:top_level} depicts the top-level architecture of our preprocessing engine for LMMSE-based data detection that implements the procedure detailed in Sec.~\ref{sec:inversionprocedure}. 
Specifically, our architecture performs: 
(i) Gram-matrix computation as in~\fref{eq:regularizedgrammatrix} using a systolic array followed by buffering the result in a flip-flop-based register array; 
(ii) matrix factorization as in Alg.~\ref{alg:blockchol} using a specialized BLDL factorization engine; and
(ii) backward substitution as in Sec.~\ref{sec:inversionprocedure} by reusing the systolic array. 
The complete schedule (in clock cycles) for a $64\times16$ channel matrix~$\bH$ is depicted in \fref{fig:mothra_schedule}. Since the matrix factorization step (ii) dominates the preprocessing latency, we utilize a BLDL-based approach, which enables higher parallelism than a Cholesky-, LDL-, LU-, or QR-based matrix-inversion approach.

\subsection{Systolic Array}\label{subsec:systolic}

\begin{figure}[tp]
\centering
\includegraphics[width=0.95\columnwidth]{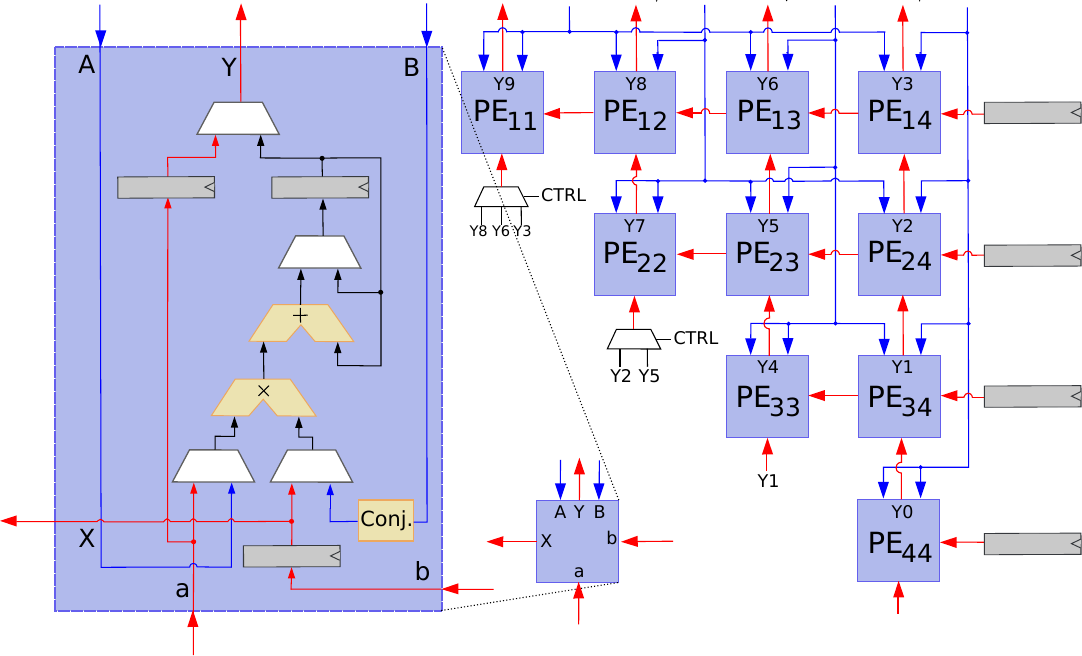}
\vspace{-0.4cm}
\caption{Architecture details of the systolic array that supports two modes: Gram-matrix computation (blue datapath) and backward substitution (red datapath). We illustrate an architecture for $U=4$ users.} \label{fig:systolic}
\vspace{-0.1cm}
\end{figure} 
As illustrated in \fref{fig:systolic}, we use a systolic array that supports two different modes. In the first mode, the matrix $\bA$ from \fref{eq:regularizedgrammatrix} is computed. 
The systolic array consists of an upper-triangular array of processing elements (PEs), each containing a complex-valued multiplier (built from four real-valued multipliers) and an accumulator. Since $\bA$ is Hermitian, we only need to compute the upper-triangular part. Since our ASIC (see Sec.~\ref{sec:asic}) is designed for a channel matrix $\bH$ of dimension $B=64$ times $U=16$, the systolic array consists of $(U^2+U)/2$ PEs. 
In every clock cycle, the systolic array is fed with the $i$th row of $\bH$, where each PE $(m,n)$ (denoted by $\textnormal{PE}_{nm}$) sequentially computes the entry $g_{mn} = \sum_{j=1}^{B}h_{mj}h_{nj}^{*}$ of $\bG=\bH^\Herm\bH$ in $B$ clock cycles. Thus, our ASIC takes $B=64$ clock cycles to calculate $\bG$.
One additional clock cycle is used to add the regularization term $\frac{N_{0}}{E_s}$ to the diagonal of $\bG$ to arrive at $\bA$. 

Subsequently, the entries of $\bA$ are buffered in the register array from (and to) which the BLDL factorization engine (see Sec.~\ref{sec:BLDLarchitecture}) can read (and write). 
The register array has $(\frac{U^2}{2}+U)/4$ entries for the  $2\times 2$ submatrices of $\bA$ and extra $U/2$ entries to store the submatrices $\bD^{-1}_{jj}$, $j=1,\ldots,\frac{U}{2}$.

In the second mode, the systolic array computes $\bA^{-1}$ using backward substitution. In the first clock cycle, the block diagonal PEs are loaded with $\bD_{jj}, j=1,\ldots,\frac{U}{2}$, submatrices computed during BLDL factorization.  
In the second clock cycle, the PE at the bottom passes the result to the PEs above, which multiply the received value with the appropriate $-l^*_{ij}$ value, accumulate the result in the internal register, and pass it to the PEs above and so on. For the $U$th column of the systolic array, this procedure is equivalent to solving the set of equations in~\fref{eq:e7}.
In the fourth clock cycle, the PE at the bottom of the $(U-1)$th column can start the same procedure described above, which corresponds to solving the set of equations in\;\fref{eq:e8}. The computations continue analogously for all the columns of the systolic array until the first column is reached. The backward-substitution step takes a total of $2U$ clock cycles.

\subsection{Block-LDL Factorization Engine}
\label{sec:BLDLarchitecture}

\begin{figure}[tp]
\centering
\includegraphics[width=0.93\columnwidth]{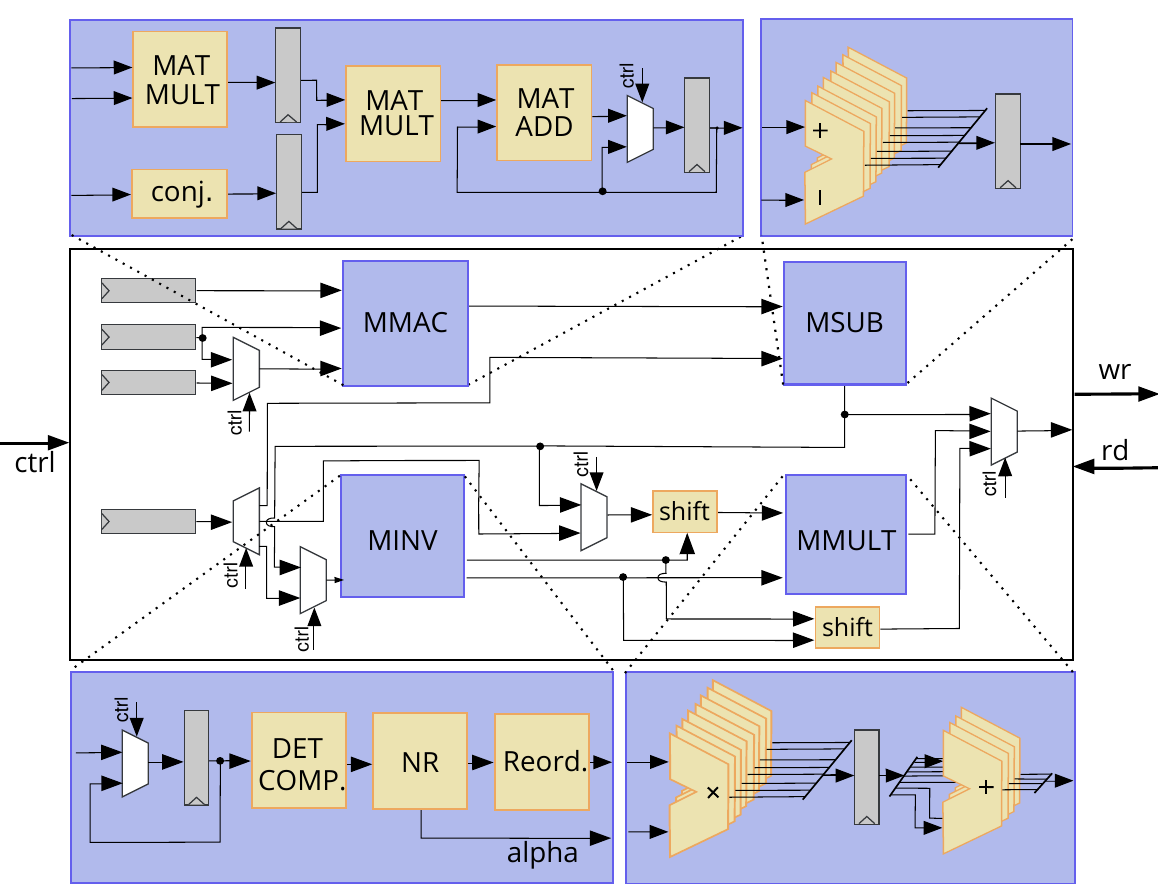}
\vspace{-0.3cm}
\caption{BLDL-factorization engine. The architecture is composed of four arithmetic units: MMAC, MSUB, MINV and MMULT. The units operate on $2\times2$ matrices and are controlled by an FSM in a processor-like fashion.}
\label{fig:ldl}
\end{figure}
Our BLDL-factorization engine implements a processor-like architecture, which is illustrated in \fref{fig:ldl}. Specialized instructions, along with data fetch and result write addresses, are encoded and stored in a look-up table (LUT). The instructions not bound to any data dependency are parallelized to minimize latency. In each clock cycle, one row of the LUT is read, triggering the state transitions of a finite-state machine (FSM), which provides the control signals to four main arithmetic units, each processing $2\times2$ submatrices: 
Matrix multiply-accumulate (MMAC), matrix subtraction unit (MSUB), a matrix inversion unit (MINV) composed of a complex-valued scalar inversion unit based on the Newton-Raphson iteration as in~\cite{studer}, and matrix-matrix multiplication (MMULT).  
The latencies of the MMAC, MSUB, MINV, and MMULT units are two, one, four, and one clock cycle(s), respectively.
\subsection{Numerical Precision }\label{sec:precision}
To optimize efficiency, we exclusively utilize fixed-point arithmetic. Before feeding the rows of the channel matrix $\bH$ to the preprocessing engine, we assume that the entries in each row are normalized by the maximum absolute value in that row. 
This enables the use of 42\,bit to represent a complex number (21\,bit per part), respectively.
To demonstrate the accuracy of our fixed-point design for a $64\times16$ massive MU-MIMO system with 16-QAM, we simulate the uncoded bit-error rate (BER). We use the QuaDRiGa mmMAGIC UMi~\cite{QuaDRiGa_tech_rpt} channel model (LoS and non-LoS) with $1^{\circ}$ minimum user separation and perfect power control, and we perform least-squares channel estimation followed by LMMSE-based data detection.

\begin{figure}[tp]
\centering
\hfill
\subfigure[non-line-of-sight (non-LoS)]{\includegraphics[width=4cm]{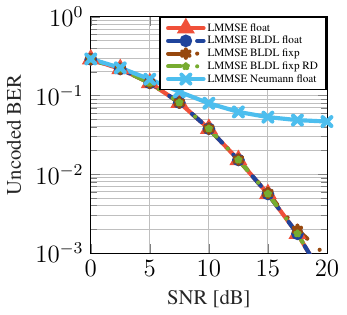}}
\hfill
\subfigure[line-of-sight (LoS)]{\includegraphics[width=4cm]{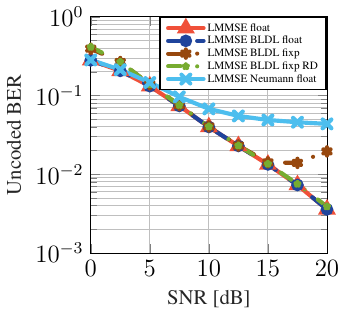}}
\hfill
\vspace{-0.15cm}
\caption{Uncoded bit-error rate (BER) vs. signal-to-noise ratio (SNR) of LMMSE-based equalization for (a) non-LoS and (b) LoS mmWave channels.}
\label{fig:bler}
\end{figure}

Fig.~\ref{fig:bler} compares the uncoded BER between a floating-point reference and our fixed-point golden model. As another baseline, we also show the performance of the approximate inversion method from~\cite{abbas}.
We also compare an alternative architecture in which~$\Delta$ in~\fref{eq:3} is forced to be real-valued (indicated by RD).
We observe that the BER of our preprocessing engine follows closely that of the floating-point reference to an uncoded BER of about $10^{-3}$ under non-LoS conditions.

%% file: 4-implementation_results.tex

\section{Implementation Results and Comparison}\label{sec:asic}

\begin{figure}[tp]
\centering
\subfigure[Photo of our Mothra chip.]{\includegraphics[width=3.5cm]{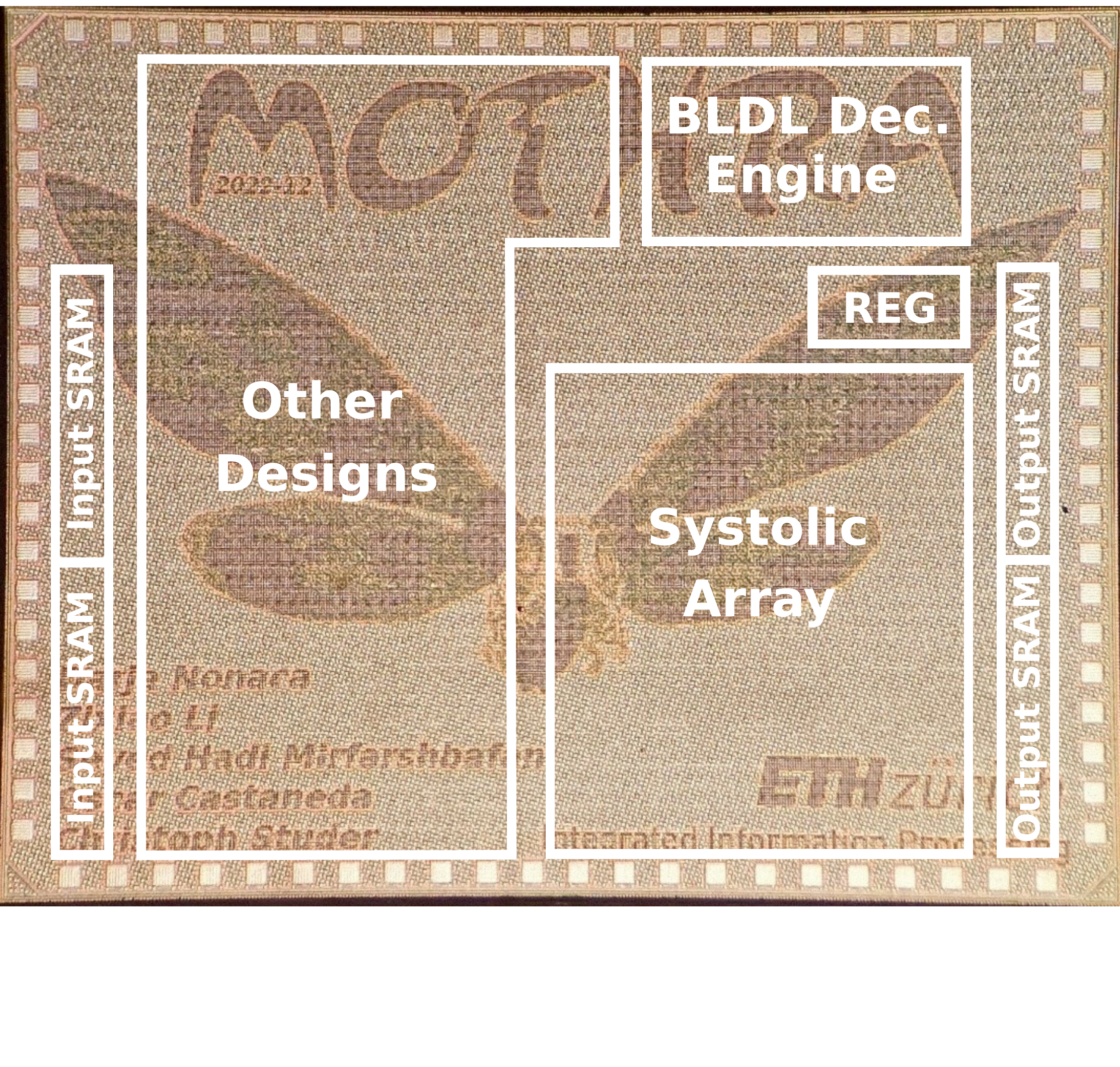}}
\subfigure[Measurement results.]{\includegraphics[width=5.1cm]{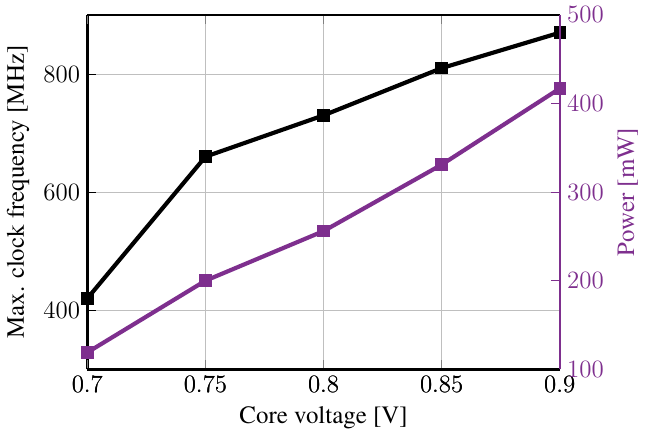}}
\vspace{-0.2cm}
\caption{(a)~Die photo of our $2.5\!\times\!2$\,mm$^\textnormal{2}$ Mothra chip containing the proposed preprocessing engine (right) along with other designs (left). (b)~Clock frequency and power measurements of the preprocessing engine.}
\label{fig:chip_photo_and_power}
\end{figure}

\fref{fig:chip_photo_and_power}(a) depicts the fabricated 5\,mm$^\textnormal{2}$ chip in GlobalFoundries' 22 FDX\textsuperscript{\sf\tiny TM} FD-SOI technology. 
The BLDL factorization engine occupies 0.2\,mm$^\textnormal{2}$, the systolic array 1\,mm$^\textnormal{2}$, and the register file 0.04\,mm$^\textnormal{2}$.
The ASIC also includes input and output SRAMs to perform high-speed measurements.
At nominal $0.9$\,V core supply and ${25}^\circ$\,C, the ASIC achieves a maximum clock frequency of $870$\,MHz with the critical path in the PE of the systolic array. 
To measure power, we create mmWave channel-matrix stimuli to loop the preprocessing engine for $7$\,ms. We do the same for accessing the input memory in order to isolate the dynamic power of the preprocessing engine by subtracting the power of such accesses from the total power. At the maximum clock frequency, the preprocessing engine consumes $416$\,mW. 
At $0.7$\,V core supply, our design achieves $420$\,MHz, which results in a power consumption of $120$\,mW and a throughput of $1.44$\,M\,mat/s and $0.7$\,\textmu{}s latency; please refer to \fref{fig:chip_photo_and_power}(b) for more details. 
\begin{table}[tp]
\setlength{\tabcolsep}{4pt} 
    \caption{Implementation results and comparison with other designs}
    \label{tab:comp}
  \vspace{-0.1cm}
    \centering
    \scalebox{0.91}{
    \begin{tabular}{@{}lccccc@{}}
        \toprule
        ~ & \multicolumn{1}{c}{This} & Mahdavi & Abbas & Kumar& Han \\
        ~ & \multicolumn{1}{c}{work}& \cite{mahdavi} & \cite{abbas} & \cite{kumar} & \cite{han} \\
        \midrule
        Algorithm   & BLDL          & Cholesky      & Neumann       & QR             & LDL \\
        $\bH$ dimension  & 64$\times$16 & 128$\times$16 &80$\times$16   & 25$\times$25   &32$\times$32\\
        Precision\;[bit]            & 21           &   20         & 16          & 32             & --\\
        Fabricated?           & yes &  no        & no          &no    &no \\
        \midrule
        Technology            & 22\,nm      &   28\,nm         & 65\,nm     & FPGA         &FPGA\\
        Core supply~[V]       & 0.9          &  --              & --         & --           & -- \\
        Active area\;[$\text{mm}^2$] & 1.204          &   --             & --         & --           & --  \\
        Cell area\;[kGE]      & 6\,030         &  537             &  117       & --           &  --\\
        Max.~clock freq.\;[MHz]  & 870          &   510        & 460         &103.8           &275\\
        Max.~throughput~[M\,mat/s]     & 1.44         &  --        & 0.54        & 5.6$\cdot$10$^\textnormal{-4}$  &3.2$\cdot$10$^\textnormal{-3}$\\
        Latency\;[\textmu{}s/mat]     & 0.7 &   --        & 1.85         &1785         &313.5\\
        Power\;[mW]            & 416          &   210        & --         &1508          & -- \\     
        \midrule
        Norm.~throughput$^\textit{a}$ \;[M\,mat/s] &   1.44       &   --         & 1.59        & --  &  --  \\
        Norm.~latency$^\textit{b}$ \;[\textmu{}s/mat] &  0.7  &   --         & 0.63& --  &  --  \\
        \bottomrule
    \end{tabular}
    } \\[0.15cm]
    \flushleft{}
    \footnotesize{\scriptsize{$^{\textit{a}}$} Scaling by $S$ and by \scriptsize{$^{\textit{b}}$} $S^{-1}$ where $S$ is the relative dimension to 22\,nm.}
\end{table}
As it can be seen from Tbl.~I, our ASIC achieves significantly higher throughput and lower latency than the FPGA designs in~\cite{kumar, han}. Our exact matrix inversion attains a comparable latency and throughput as \cite{abbas}, which achieves poor BER performance (cf.~\fref{fig:bler}). 
It is challenging to compare our design to \cite{mahdavi} as (i) they only provide synthesis results and (ii) the preprocessing latency was not reported. Furthermore, their reported cell area appears to be unusually compact, especially when considering that matrices of dimension $128\times16$ are processed.

%% file: 5-conclusion.tex
\section{Conclusions}\label{sec:conclusion}
We have proposed the first fabricated and measured preprocessing ASIC for LMMSE-based data detection in a $64$ BS antenna, $16$ user mmWave massive MU-MIMO system.
Unlike existing matrix-factorization approaches, our BLDL-based design improves parallel processing, thereby reducing preprocessing latency. Our ASIC achieves a throughput of $1.44$ M\,mat/s and $416$\,mW at $870$\,MHz clock frequency at a latency of only $0.7$\,\textmu{}s.
When compared to existing preprocessing engines, our implementation outperforms other designs in terms of throughput, area, latency, and/or error-rate performance. 
%

%% file: 24ESSERC_MOTHRA.bbl
\begin{thebibliography}{10}
\providecommand{\url}[1]{#1}
\csname url@samestyle\endcsname
\providecommand{\newblock}{\relax}
\providecommand{\bibinfo}[2]{#2}
\providecommand{\BIBentrySTDinterwordspacing}{\spaceskip=0pt\relax}
\providecommand{\BIBentryALTinterwordstretchfactor}{4}
\providecommand{\BIBentryALTinterwordspacing}{\spaceskip=\fontdimen2\font plus
\BIBentryALTinterwordstretchfactor\fontdimen3\font minus
  \fontdimen4\font\relax}
\providecommand{\BIBforeignlanguage}[2]{{%
\expandafter\ifx\csname l@#1\endcsname\relax
\typeout{** WARNING: IEEEtran.bst: No hyphenation pattern has been}%
\typeout{** loaded for the language `#1'. Using the pattern for}%
\typeout{** the default language instead.}%
\else
\language=\csname l@#1\endcsname
\fi
#2}}
\providecommand{\BIBdecl}{\relax}
\BIBdecl

\bibitem{busari}
S.~A. Busari \emph{et~al.}, ``Millimeter-wave massive {MIMO} communication for
  future wireless systems: A survey,'' \emph{{IEEE} Comm. Surveys \&
  Tutorials}, 2018.

\bibitem{juntti}
M.~A. Albreem \emph{et~al.}, ``Massive {MIMO} detection techniques: A survey,''
  \emph{{IEEE} Comm. Surveys \& Tutorials}, 2019.

\bibitem{mwu}
M.~Wu \emph{et~al.}, ``Implicit vs. explicit approximate matrix inversion for
  wideband massive {MU-MIMO} data detection,'' in \emph{{IEEE} JSPS}, 2018.

\bibitem{clasen}
R.~Clasen, \emph{Numerical methods for inverting positive definite
  matrices}.\hskip 1em plus 0.5em minus 0.4em\relax {RAND} Corporate Tech.
  Rep., 1966.

\bibitem{golub}
G.~H. Golub \emph{et~al.}, \emph{Matrix Computations}, 3rd~ed.\hskip 1em plus
  0.5em minus 0.4em\relax The Johns Hopkins Univ. Press, 1996.

\bibitem{burg06}
A.~Burg \emph{et~al.}, ``Algorithm and {VLSI} architecture for linear {MMSE}
  detection in {MIMO-OFDM} systems,'' in \emph{IEEE ISCAS}, 2006.

\bibitem{studer}
C.~Studer \emph{et~al.}, ``{ASIC} implementation of soft-input soft-output
  {MIMO} detection using {MMSE} parallel interference cancellation,''
  \emph{{IEEE} JSSC}, 2011.

\bibitem{auras}
D.~Auras \emph{et~al.}, ``Efficient {VLSI} architectures for matrix inversion
  in soft-input soft-output {MMSE} {MIMO} detectors,'' in \emph{IEEE ISCAS},
  2014.

\bibitem{omran}
S.~S. Omran \emph{et~al.}, ``Fast {QR} decomposition based on {FPGA},'' in
  \emph{IEEE ICOASE}, 2018.

\bibitem{singh}
C.~K. Singh \emph{et~al.}, ``{VLSI} architecture for matrix inversion using
  modified {G}ram-{S}chmidt based {QR} decomposition,'' in \emph{VLSID}, 2007.

\bibitem{mahdavi}
M.~Mahdavi \emph{et~al.}, ``A {VLSI} implementation of angular-domain massive
  {MIMO} detection,'' in \emph{IEEE ISCAS}, 2019.

\bibitem{abbas}
S.~Abbas \emph{et~al.}, ``Low-latency approximate matrix inversion for
  high-throughput linear pre-coders in massive {MIMO},'' in \emph{IFIP/IEEE
  VLSI-SoC}, 2016.

\bibitem{prabhu}
H.~Prabhu \emph{et~al.}, ``Hardware efficient approximative matrix inversion
  for linear pre-coding in massive {MIMO},'' in \emph{IEEE ISCAS}, 2014.

\bibitem{stan}
I.~{Stanimirovic}, ``Full-rank block {$LDL^*$} decomposition and the inverses
  of nxn block matrices,'' in \emph{JAMC}, 2012.

\bibitem{QuaDRiGa_tech_rpt}
S.~Jaeckel \emph{et~al.}, ``{QuaDRiGa} - quasi deterministic radio channel
  generator user manual and documentation,'' Fraunhofer Heinrich Hertz
  Institute, Tech. Rep. v2.0.0, Aug. 2017.

\bibitem{kumar}
K.~V.~S. Kumar \emph{et~al.}, ``System on chip implementation of floating point
  matrix inversion using modified {G}ram-{S}chmidt based {QR} decomposition on
  {PYNQ} {FPGA},'' in \emph{IEEE iSES}, 2021.

\bibitem{han}
K.~Han \emph{et~al.}, ``Low-latency {FPGA} design and implementation of
  {H}ermitian matrix inversion based on partitioned systolic array for massive
  {MIMO},'' in \emph{IEEE ICTA}, 2022.

\end{thebibliography}
